\documentclass[sigconf]{acmart}

\usepackage{balance}

\def\BibTeX{{\rm B\kern-.05em{\sc i\kern-.025em b}\kern-.08emT\kern-.1667em\lower.7ex\hbox{E}\kern-.125emX}}

\usepackage{libertine}

\usepackage{booktabs} 
\usepackage{mathtools}
\usepackage{subcaption}
\usepackage{caption}

\usepackage{algorithm}
\usepackage[noend]{algpseudocode}

\usepackage{mathrsfs} 
\usepackage{amsmath}
\usepackage{amsbsy}

\usepackage{amsmath,amssymb,amsfonts}
\usepackage{textcomp}
\usepackage{xcolor}

\usepackage{booktabs} 
\usepackage{mathtools}
\usepackage{subcaption}

\usepackage{import}
\usepackage{graphicx}
\graphicspath{ {./images/} }

\copyrightyear{2020}
\acmYear{2020}
\setcopyright{acmcopyright}
\acmConference[WSDM '20]{The Thirteenth ACM International Conference on Web Search and Data Mining}{February 3--7, 2020}{Houston, TX, USA}
\acmBooktitle{The Thirteenth ACM International Conference on Web Search and Data Mining (WSDM '20), February 3--7, 2020, Houston, TX, USA}
\acmPrice{15.00}
\acmDOI{10.1145/3336191.3371814}
\acmISBN{978-1-4503-6822-3/20/02}

\acmSubmissionID{wsdm489}

\begin{document}

\fancyhead{}

\title{Listwise Learning to Rank by Exploring Unique Ratings}


\author{Xiaofeng Zhu}
\email{xiaofengzhu2013@u.northwestern.edu}
\affiliation{%
  \institution{Northwestern University}
  \city{Evanston}
  \state{IL}
  \country{USA}}

\author{Diego Klabjan}
\email{d-klabjan@northwestern.edu}
\affiliation{%
  \institution{Northwestern University}
  \city{Evanston}
  \state{IL}
  \country{USA}}

%
\renewcommand{\shortauthors}{Zhu and Klabjan}

%
\begin{abstract}
In this paper, we propose new listwise learning-to-rank models that mitigate the shortcomings of existing ones. Existing listwise learning-to-rank models are generally derived from the classical Plackett-Luce model, which has three major limitations. (1) Its permutation probabilities overlook ties, i.e., a situation when more than one document has the same rating with respect to a query. This can lead to imprecise permutation probabilities and inefficient training because of selecting documents one by one. (2) It does not favor documents having high relevance. (3) It has a loose assumption that sampling documents at different steps is independent. To overcome the first two limitations, we model ranking as selecting documents from a candidate set based on unique rating levels in decreasing order. The number of steps in training is determined by the number of unique rating levels. More specifically, in each step, we apply multiple multi-class classification tasks to a document candidate set and choose all documents that have the highest rating from the document set. This is in contrast to taking one document step by step in the classical Plackett-Luce model. Afterward, we remove all of the selected documents from the document set and repeat until the remaining documents all have the lowest rating. We propose a new loss function and associated four models for the entire sequence of weighted classification tasks by assigning high weights to the selected documents with high ratings for optimizing Normalized Discounted Cumulative Gain (NDCG). To overcome the final limitation, we further propose a novel and efficient way of refining prediction scores by combining an adapted Vanilla Recurrent Neural Network (RNN) model with pooling given selected documents at previous steps. We encode all of the documents already selected by an RNN model. In a single step, we rank all of the documents with the same ratings using the last cell of the RNN multiple times. We have implemented our models using three settings: neural networks, neural networks with gradient boosting, and regression trees with gradient boosting. We have conducted experiments on four public datasets. The experiments demonstrate that the models notably outperform state-of-the-art learning-to-rank models.
\end{abstract}


%

%

%
\maketitle

\section{Introduction}

Learning-to-rank is one of the most classical research topics in information retrieval, and researchers have put tremendous efforts into modeling ranking behaviors. In training, existing ranking models learn a scoring function from query-document features and multi-level ratings/labels, e.g., 0, 1, 2. During inference, the learned scoring function is used to obtain prediction scores for ordering documents. There are two major challenges for learning-to-rank. The first challenge is that there can be a mismatch between ratings and the correct ranking orders in training. Although it is straightforward to infer partial ranking orders from ratings and prediction scores, it is not easy to design loss functions modeling the order of ratings and the order of prediction scores. Many prediction scores indicate the same ranking, i.e., the order of documents. This implies that a model does not necessarily have to match ratings, opening opportunities and ambiguities. Moreover, the top ranking positions are more important. The second challenge is that raw features may not be representative enough for learning a reasonable scoring function. Existing ranking models tackle the two challenges by: (1) designing loss functions or reward functions to map prediction scores with correct ranking orders in training, and (2) tuning loss functions with evaluation metrics such as NDCG \cite{jarvelin2002cumulated}, or ERR \cite{chapelle2009expected}, and (3) calculating prediction scores using richer features such as a local ranking context \cite{pang2017deeprank, ai2018dlcm, bello2018seq2slate, albuquerque2018learning, guo2019deep}. 

Depending on how prediction scores in training are compared with ratings, there are three types of loss functions: pointwise, pairwise, and listwise. Pointwise learning maps the prediction scores of individual documents with their exact ratings \cite{liu2009learning}, which is not necessary for obtaining correct orders. Pairwise learning \cite{herbrich1999large, freund2003efficient, burges2005learning, li2007ordinal, quoc2007learning, zhou2008learning, sculley2009large, burges2010ranknet, wang2018lambdaloss} naturally compares pairs of documents to minimize the number of inversions. Earlier pairwise models such as RankSVM \cite{herbrich1999large}, RankBoost \cite{freund2003efficient}, RankNet \cite{burges2005learning}, and Ordinal Regression \cite{li2007ordinal} may overly update weights for different pairs as they treat all pairs with equal importance. For instance, suppose the correct order is (a -> b -> c -> d). When a pairwise model catches an inversion (c -> b) in a predicted order (a -> c -> b -> d), it tries to update weights to increase the score of b and decrease the score of c. However, if the score of b becomes too high -- higher than a -- this causes another inversion (b, a). LambdaMart \cite{quoc2007learning, burges2010ranknet, hu2019unbiased} and NDCG-LOSS++ \cite{wang2018lambdaloss} largely limit this issue by assigning different weights for different pairs when calculating their gradients \cite{burges2010ranknet}. Their best models rely on using gradient boosting regression trees \cite{chen2016xgboost, meng2016communication, ke2017lightgbm}, which are effective but very sensitive to hyper-parameters. Listwise learning \cite{cao2007learning, xia2008listwise, taylor2008softrank, guiver2009bayesian, niu2012new, lan2014position, luo2015stochastic, jiang2018beyond} tries to learn the best document permutation based on permutation probabilities proposed by Plackett \cite{plackett1975analysis} and Luce \cite{luce2012individual}. The classical Plackett-Luce model has a constraint that the permutation probabilities are targeted for unique ratings. For instance, maximizing the likelihood of choosing a document from a set of documents that have the same rating can confuse a model. Since the number of unique rating levels is typically much smaller than the number of candidate documents per query, there can be a large number of ties. Also, in order to calculate the joint probability of a ranking sequence, the number of steps a Plackett-Luce model, such as ListMLE \cite{xia2008listwise}, needs to go through is bound by the number of candidate documents per query. Therefore, obtaining one permutation can be computationally inefficient. Furthermore, top documents are more important, but each step of a Plackett-Luce model is equally important along the entire sequence. Variants such as SoftRank \cite{taylor2008softrank}, p-ListMLE \cite{lan2014position}, and ApproxNDCG \cite{bruch2019revisiting} use NDCG or ranking positions to tune their loss functions. Nevertheless, when the number of documents in a permutation is large, gradients can vanish or explode very fast as the product of their permutation probabilities is close to 0. Highlights from research studies in recent years for scoring functions include ensemble scoring functions \cite{liu2009learning, ai2018learninggroupwise, Gallagher:2019:JOC:3289600.3290986}, ranking refinement \cite{ai2018dlcm}, and reinforcement learning \cite{wei2017reinforcement, feng2018greedy, liu2018novel, zeng2018multi, oosterhuis2018ranking, luo2018dynamic}. Despite being effective, training efficiency is still their common bottleneck, which limits their usage in real-world applications.

We propose a new listwise loss function and associated four models to address the issues of existing ranking models. Existing Plackett-Luce models use $n-1$ steps, and $n$ is the number of documents in a query. In contrast, our models learn to gradually select documents in descending order of their ratings using $|R| - 1$ steps, and $R$ denotes the set of their unique ratings. The number of unique ratings $|R|$ is usually much smaller than $n$ in real-world ranking data. Each step corresponds to a ranking level and an associated candidate document set. There are $n$ documents in the candidate document set at the first step. At each step, there can be multiple documents that have the highest rating in the associated candidate document set. For each such highest rating document, we apply a softmax to the document and all of the documents that have lower ratings in the current document set. We also set a weight based on the current highest rating for this step to favor top ranking levels. We select all of the highest rating documents for the current step, then remove them from the document set to move to the next step. The models embed all $n$ documents and produce directly $n$ scores based on neural networks, neural networks with gradient boosting, or gradient boosting regression trees. To tackle the third limitation of Plackett-Luce models, we adapt RNN functions and pooling to the second model for learning document-query relevancy given selected documents, which we call conditional prediction scores. The RNN functions learn transitions from documents that have the highest rating at one step to documents that have the highest rating at the next step.

The major contributions of our work are as follows. (1) We solve the multi-level ranking problem by selecting documents from the highest ranking level to the lowest ranking level. In each step, we conduct multiple weighted multi-class classification tasks, which are eventually transferred to likelihood maximization tasks. (2) We propose a new loss function that handles ties in ranking data and optimizes NDCG. (3) We also propose a new way of utilizing adapted RNN functions for obtaining conditional prediction scores in training.


\section{Background}
\label{Background}
In this section, we briefly review the most related previous works and address their limitations in detail.

\noindent \textbf{ListNet}
Since learning the complete $n!$ permutations is intractable, ListNet \cite{cao2007learning} generally minimizes the cross-entropy of top-one probabilities of prediction scores and ratings using a softmax function. Given a set of $n$ documents for a specific query $D = \{d_i\}_i$, their ratings $Y = \{y_i\}_i$, and a global scoring function $f$, the loss function for ListNet using top-one probabilities is
$$L(f; D, Y) = - \sum_{i=1}^n P_i(Y) log \ P_i(\{f(d_i)\}),$$

\noindent where $P_i(Y) = \frac{exp(Y_i)}{\sum_{j=1}^n exp(Y_j)}$.
 
Training ListNet using top-one probabilities is fast, but its critical issue is that minimizing loss is not always consistent with the correct ranking order. This is due to the fact that the true probabilities $\{P_i(Y)\}_i$ are defined on a softmax function of ratings. For instance, given ratings (2, 1), a correct order with prediction scores (20, 1) has a higher ListNet loss compared to a wrong order with prediction scores (1, 2), because the top-one probabilities of (1, 2) is closer to the top-one probabilities of the ratings.

\noindent \textbf{ListMLE}
The other type of listwise learning, ListMLE, only maximizes the likelihood of the permutation probabilities of one permutation corresponding to a correct ranking. The loss function for ListMLE is

$$L(f; D) = -log \ P(D|f), P(D |f) = \prod_{i=1}^n P_i (\{\tilde{d_k}\}_{k=i}^n),$$

\noindent where $\tilde{d}_i$ is the document ranked at position $i$.

A correct ranking sequence satisfies the property that for any two documents $\tilde{d}_i$ and $\tilde{d}_j$, if the rating of $\tilde{d}_i$ is higher than the rating $\tilde{d}_j$, $\tilde{d}_i$ is ranked before $\tilde{d}_j$ in the ordered sequence. Since there can be ties in ratings, a sampling method of selecting a correct sequence is generally used \cite{jagerman2017modeling}. A ListMLE model does not have the cross-entropy issue but can suffer from sampling $n!$ correct orders when all documents have the same rating. More importantly, when $n$ is large the likelihoods at the top positions become very small, which leads to imbalance gradient updates for documents ranked at different positions.

\noindent\textbf{RNN}
The Vanilla RNN model takes a sequence of feature vectors $X=(x_t)_t$ as input, where $x_t$ indicates the feature vector of the document at step $t$, and computes the hidden output $h_t$ at each step. We use $h_t = rnn(x_t, h_{t-1}, W)$ in this paper to represent an RNN function, where $W$ is a set of trainable weight matrices. Although RNNs can learn the conditional/hidden transitions, it is intractable to apply an RNN to the complete order of documents for a query because of the following two reasons. (1) Some documents have the same rating due to ties. (2) Training a long sequence can be time and memory consuming. It still easily suffers gradient vanishing even if it utilizes more advanced RNN models such as a Long Short-term Memory network (LSTM) \cite{hochreiter1997long} or a Gated Recurrent Unit network (GRU) \cite{chung2014empirical}. A common practice of applying RNNs in learning-to-rank is by refining the top positions of a ranked list using two training phases \cite{ai2018dlcm}. The novelty of our RNN-based model is that we apply RNN and pooling functions to multiple documents at each step.

\section{Proposed Loss Function and Models}
\label{Proposed Loss Function and Models}
Suppose each query $q$ in a training data set is associated with a set of $n^{(q)}$ candidate documents. Each document has query-document features and a rating. We derive a scoring function $f$ by minimizing the average loss between the scores estimated for candidate documents and their ratings. For convenience, we drop the dependency of $q$ for the rest of this paper. 
Different queries may have a different number of documents, however, this does not affect our model. We next present the loss function and the associated four models for scoring function $f$.

Given the set of documents $D$ for a query, look-up function $r(d)$ denotes the rating of document $d \in D$, and scoring function $f(d)$ returns the prediction score of document $d$. Let the sorted unique ratings of this query in descending order be $R = \{r_1, r_2,  ..., r_{|R|}\}$, $r_1 > r_2 > ... > r_t > ... > r_{|R|}$, where $r_t$ denotes the $t^{th}$ highest rating. We model ordering the $n$ documents as selecting the most relevant documents using $|R|$ steps. At each step, we select all documents that have the highest rating from a candidate document set for this step. We use $S = \{s_1, s_2, ..., s_{|R|}\}$ to hold the candidate document sets at different steps. The initial set $s_1$ is $D$.

Let $c_t$ denote the set of documents whose ratings are equal to $r_t$ from the document candidate set $s_t$ and let  $\tilde{s}_t$ denote the set of documents having ratings lower than $r_t$. We have $s_t = c_t \cup \tilde{s}_t$. At step $t$, we learn to identify all documents in $c_t$ from $s_t$. A straightforward approach of applying a softmax to $s_t$ would yield the same cross-entropy issue outlined in Section \ref{Background}. More specifically, if we encode class labels using 0s and 1s, where 1 indicates a positive class, i.e., a document being in $c_t$, and 0 indicates a negative class, the class labels can have multiple 1s. The novel idea of our approach is to have several softmax operations with each one targeting a single document from $c_t$. To this end, we generate several sets $\bar{s_t}^k = \tilde{s}_t \cup \{d_k\}$ for each $d_k \in c_t$. Each smaller set contains only one document that has rating $r_t$ and the rest have ratings lower than $r_t$. By doing so, the class labels of each multi-class classification task only have one 1 and the rest are 0s. Since cross-entropy is equivalent to likelihood maximization for Dirac distributions ($\bar{s_t}^k$ can be interpreted as such since only $d_k$ has "label 1"), we can therefore break ties by maximizing the likelihood of selecting each document that has the highest rating $r_t$ among itself and documents that have lower ratings. The joint probability of selecting all documents in $c_t$ is approximated by the product of their likelihoods
$$P(c_t | s_t) \approx \prod_{d_k \in c_t} P(d_k | \bar{s_t}^k).$$

We then model $P(d_k | \bar{s}_t^k)$ through the usual softmax expression and use the log likelihood

$$ln \ P(c_t | s_t) \approx \sum_{d_k \in c_t} ln \ P(d_k | \bar{s_t}^k).$$

For instance, suppose there are four documents $\{a, b, c, d\}$, and their ratings are $\{3, 3, 2, 1\}$. In order to hold the correct ranking order $\{a, b, c, d\}$ or $\{b, a, c, d\}$, document $a$ only needs to outrank documents $c$ and $d$, document $b$ only needs to outrank documents $c$ and $d$, and document $c$ only needs to outrank document $d$; the order of documents $a$ and $b$ does not influence the performance of metrics such as NDCG. We generate two sets $\{a, c, d\}$ and $\{b, c, d\}$ from the original set and encode their class labels $\{1, 0, 0\}$ and $\{1, 0, 0\}$. We select both documents $a$ and $b$ by maximizing the likelihood of selecting $a$ among $\{a, c, d\}$ and maximizing the likelihood of selecting $b$ from $\{b, c, d\}$. The joint probability of selecting both $a$ and $b$ from the original set $\{a, b, c, d\}$ is approximated by the product of their likelihoods. Similarly, we further generate set $\{c, d\}$ and encode the class labels as $\{1, 0\}$ and maximize the likelihood of selecting $c$ from $\{c, d\}$. 

In the end, we remove those documents that have the highest rating from $s_t$ to generate the next set $s_{t+1}$ and repeat until we reach set $s_{|R|}$. More specifically, $s_{t+1} = s_{t} \setminus c_t = \tilde{s}_t$. 

The likelihood of selecting a document $d \in c_t$ is defined as

\begin{equation} \label{P_t(d)}
P_t(d) = \frac{exp(f(d))}{exp(f(d)) + \sum_{d' \in \tilde{s}_t} exp(f(d'))}.
\end{equation}

\noindent Recall that the rating of every document in $\tilde{s}_t$  is less than $r_t$.

The loss function is defined as
\begin{equation} \label{loss}
L(f; R, D) = -\frac{1}{|R| - 1} \sum_{t=1}^{|R| - 1} (2^{r_t} - 1) \sum_{d \in c_t} ln \  P_t(d),
\end{equation}

\noindent where $|R| > 1$, and $2^{r_t} - 1$ is used to boost the importance of relevant documents.

Suppose there are four unsorted documents $D$ for a query, their ratings are $\{1, 2, 2, 0\}$, the sorted unique ratings are $R = \{2, 1, 0\}$, and their prediction scores returned from $f$ are $\{ln \ 2, ln \ 3, ln \ 4, ln \ 5\}$. The loss is calculated as
\begin{align*}
L(f; R, D) &= - \frac{1}{2} \{(2^2 - 1) (ln \ \frac{3}{3 + 2 + 5} + ln \ \frac{4}{4 + 2 + 5}) \\
& + (2^1 -1) (ln \ \frac{2}{2 + 5})\}.
\end{align*}

When all $n$ documents have the same rating, we do not need a ranking function for sorting the documents, and thus the loss is 0. Similarly, at step $|R|$ all documents in set $s_{|R|}$ have the same rating. Therefore we only consider loss from steps 1 to $|R| - 1$. This avoids processing documents in $s_{|R|}$ and can further make training efficient because the click-through-rate of real-world ranking data is usually low, i.e., $\frac{|s_{|R|}|}{|D|}$ is high. If $|R| = n$, i.e., there are no ties, $P_t(d)$ becomes the likelihood function addressed in ListMLE, however, in real-world ranking data $|R| << n$. 

Since the loss function is essentially based on maximum likelihood estimation, when the number of documents is large, the likelihood of selecting each document, especially at a top rating position, can still become small \cite{barber2016dealing}. Therefore, we further partition the set $\tilde{s}_t = \{\tilde{s}_{t}^{b}\}_{b}$ in (\ref{P_t(d)}) into several smaller sets of documents. Each set of a fixed window size $u$ is based on the decreasing order of prediction scores in $\tilde{s}_t$. The likelihood is modified to
\begin{equation} \label{window_P_t(d)}
P_t(d) \approx \prod_{b} \frac{exp(f(d))}{exp(f(d)) + \sum_{d' \in \tilde{s}_{t}^{b}} exp(f(d'))}.
\end{equation}
This loss function is non-differentiable since we sort with respect to the prediction scores. The standard practice of selecting the order in the forward pass and then differentiating with respect to this fixed order in the backward pass is applicable.

In Section \ref{uRank}, we explain our neural network model uRank built upon this loss function. Based on uRank, we further propose three ensemble models uBoost, uMart, and urBoost in Sections \ref{uBoost}-\ref{urBoost} using gradient boosting, gradient boosting trees, and RNNs with gradient boosting, respectively.

\subsection{uRank Model}
\label{uRank}
Our uRank model uses the proposed loss function and a neural network with two hidden layers as a scoring function (there is no constraint on the number of layers). Let $X$ be the matrix with rows corresponding to the feature vectors of each document (combined with query features). If there are $l$ features, then $X \in \mathbb{R}^{n \times l}$. In our experiments, we utilize 3 trainable matrices $W_1 \in \mathbb{R}^{l \times k_1}$, $W_2 \in \mathbb{R}^{k_1 \times k_2}$, and $W_3 \in \mathbb{R}^{k_2 \times 1}$ for the input layer and the two hidden layers respectively to get the prediction scores as $f = \phi(\phi(\phi(X W_1)W_2) W_3) \in \mathbb{R}^{n}$, where $\phi$ is an activation function. 

We have designed an efficient tensor-based algorithm for our loss calculation inspired by the mini-batch approach for RankNet \cite{quoc2007learning}. We take the labels and prediction scores of all documents for a query as input and calculate the loss without using any loops, which can speed up training on tensor-based frameworks, such as Tensorflow, as loops are time-consuming in such frameworks. Algorithm \ref{Tensor-based loss calculation} explains the loss calculation process of one query. The inputs labels $Y$ and prediction scores $F$ are two $n$ dimensional vectors, where $n$ is the number of documents in the query. Matrix $P$ calculates the differences of prediction scores of different pairs, and it is used to create mask $M$. Matrices $P$ and $M$ are $n \times n$ dimensional. Vector $T$ calculates $\sum_{d' \in \tilde{s}_t} exp(f(d'))$ in (\ref{P_t(d)}). The operation $\cdot$ indicates a matrix multiplication, $/$ is element-wise, and $\odot$ is the Hadamard product. Also we denote $e=(1,1,\dots,1)\in \mathbb{R}^n$.

The calculation with regard to each query for our loss function is bound by $(|R| -1) \times n$ softmax terms. In contrast, the same time complexity of MLE-based models is $O(n log n)$ when candidate queries are sorted by ranking levels during training. Training uRank is faster than MLE-based models in practice because of ties. 
\begin{algorithm}
\caption{Tensor-based loss calculation}\label{Tensor-based loss calculation}
\begin{algorithmic}[1]
\State \textbf{Input}: labels $Y=(Y_i)_{i=1}^n$
\State \textbf{Input}: prediction scores $F=(f(d_i))_{i=1}^n$
\State Gains $G\gets (2^{Y_i} - 1)_{i=1}^n$
\State $P \gets Ye^T - eY^T$
\State Mask $M \gets [1$ if $v > 0, 0$ if $v \leq 0$ for $v$ in $P]$
\State $T \gets M \cdot exp(F)$
\State $L \gets ln( 1 + T / exp(F))$
\State $L \gets L \odot G$ 
\State $loss \gets \frac{sum(L)}{|R| - 1}$
\State \textbf{Return} $loss$
\end{algorithmic}
\end{algorithm}

\subsection{uBoost Model}
\label{uBoost}

uBoost is an ensemble model that combines uRank with gradient boosting based on neural networks. uBoost aims to take $M$ sequential and additive steps to find a scoring function $F_M(D)$ that best maps prediction scores and ratings \cite{friedman2001greedy}. At each step $m \leq M$, we calculate "residuals" $\{\tilde{y}_d^{(m)}\}$, optimize weights for weak-learner $f_m(D)$, and choose a model coefficient $\rho_m$ for the weak-learner. Value $\tilde{y}_d^{(m)}$ is defined as the negative gradient of loss $L(F_m; R, D)$ with respect to $F_m(d)$, where we have $f = F_M(D) = \sum_{m=1}^M \rho_m f_m(D)$. The first weak learner in uBoost is uRank. Each of the remaining weak learners is also a neural network with two hidden layers but with Mean Squared Error (MSE) as its loss function. This is the standard loss function used in the gradient boosting steps matching $\tilde{y}_d^{(m)}$ with residuals based on $P_i(Y)$ for $i$ corresponding to $d$ \cite{friedman2001greedy}.

The derivation of the residuals of a query for the $m^{th}$ ensemble model $F_m$ is calculated as
\begin{align*}
&- \frac{\partial L(F_m; R, \{\bar{d}\})}{\partial F_m(\bar{d})}\\
&= - \sum_d \frac{\partial L(F_m; R, d)}{\partial P_t(d)} \frac{\partial P_t(d)}{\partial F_m(\bar{d})},\\
&= \frac{1}{|R| - 1} \sum_{t=1}^{|R| - 1} (2^{r_t} - 1) \sum_{d \in c_t} \frac{1}{P_t(d)} \frac{\partial P_t(d)}{\partial F_m(\bar{d})}\\
&:= \sum_d \tilde{y}_d^{(m)}.\\
\end{align*}

To simplify the derivation of $\frac{\partial P_t(d)}{\partial F_m(d)}$ and $\frac{\partial P_t(d)}{\partial F_m(\bar{d})},d\ne \bar{d}$, let $k(s) = \sum_{d' \in s} exp(F_m(d'))$. We assume $P_t(d)$ as defined in (\ref{P_t(d)}), the estimation for (\ref{window_P_t(d)}) is a straightforward extension. Quantity $\frac{\partial P_t(d)}{\partial F_m(d)}$ reads

$$\frac{\partial P_t(d)}{\partial F_m(d)} = P_t(d) \frac{1}{1 + \frac{e^{F_m(d)}}{k(\tilde{s}_{r(d)})}},$$

while $\frac{\partial P_t(d)}{\partial F_m(\bar{d})},d\ne \bar{d}$ is 
$$\frac{\partial P_t(d)}{\partial F_m(\bar{d})} = -P_t(d) \frac{e^{F_m(\bar{d})}}{k(\tilde{s}_{r(d)}) + e^{F_m(d)}}.$$

The derivation of residual of $d$ reads
\begin{equation} \label{gradients_full}
\begin{split}
\tilde{y}_d^{(m)} &=\frac{1}{|R| - 1} (2^{r(d)} - 1) \frac{1}{1 + \frac{e^{F_m(d)}}{k(\tilde{s}_{r(d)})}}\\
		  &-\frac{1}{|R| - 1} \sum_{\hat{d} \in D\setminus s_{r(d)}} (2^{r(\hat{d})} - 1) \frac{e^{F_m(d)}}{k(\tilde{s}_{r(\hat{d})}) + e^{F_m(\hat{d})}}.\\
\end{split}
\end{equation}

Algorithm \ref{Tensor-based residual calculation} explains the residual calculation process and its MSE loss calculation for the $m^{th}$ weak learner $f_m(D)$ of one query.

\begin{algorithm}
\caption{Tensor-based residual calculation}\label{Tensor-based residual calculation}
\begin{algorithmic}[1]
\State \textbf{Input}: labels $Y=(Y_i)_{i=1}^n$
\State \textbf{Input}: prediction scores $F_m=(F_m(d_i))_{i=1}^n$
\State \textbf{Input}: prediction scores for residuals $(f_m(d_i))_{i=1}^n$
\State Gains $G\gets (2^{Y_i} - 1)_{i=1}^n$
\State $P \gets Ye^T - eY^T$
\State Mask $M \gets [1$ if $v > 0, 0$ if $v \leq 0$ for $v$ in $P]$
\State $T \gets M \cdot exp(F_m)$
\State $Z \gets 1 / ( 1 + exp(F_m) / T)$
\State $Z \gets Z \odot G$
\State Mask $M' \gets [-1$ if $v < 0, 0$ if $v \geq 0$ for $v$ in $P]$
\State $T' \gets T + exp(F_m)$
\State $Z' \gets M' \cdot (G / T')$
\State $Z' \gets Z' \odot exp(F_m)$
\State $Z_1 \gets Z + Z'$
\State $Z_1 \gets \frac{Z_1}{|R| - 1}$
\State $loss \gets MSE(Z_1, (f_m(d_i))_{i=1}^n)$
\State \textbf{Return} $loss$
\end{algorithmic}
\end{algorithm}

\subsection{uMart Model}
\label{uMart}

uMart differs from uRank and uBoost in that the weak learners in uMart are regression trees instead of neural networks. uMart adapts the gradient weighting factor $\Delta Z$ from the implementation of LambdaMart in LightGBM \cite{burges2005learning, meng2016communication, ke2017lightgbm}. The residual calculation and the MSE loss calculation are similar to Algorithm \ref{Tensor-based residual calculation}. uMart requires the second derivative of the loss with respect to the scoring function $F_m$ for leaf splits \cite{burges2010ranknet}. Note that LightGBM tunes $\Delta Z$ from LambdaRank \cite{burges2005learning} and the weight for the second derivative for better performance. The contribution document $d$ makes to the gradient is
\begin{equation} \label{first_derivative}
\begin{split}
&\frac{\partial L(F_m; R, \{d\})}{\partial F_m} = \sigma \sum_{d' \in \tilde{s}_{r(d)}} \Delta Z(d, d') \frac{e^{\sigma \cdot F_m(d')}}{e^{\sigma \cdot F_m(d)} + k(\tilde{s}_{r(d)})} \\
&- \sigma \sum_{\hat{d} \in D \setminus s_{r(d)}} \Delta Z(\hat{d},d) \frac{e^{\sigma \cdot F_m(d)}}{k(\tilde{s}_{r(\hat{d})}) + e^{\sigma \cdot F_m(\hat{d})}},\\
&\Delta Z(d, d') = \frac{(2^{r(d)} - 2^{r(d')}) {(discount(d) - discount(d')})}{|F_m(d) - F_m(d')| \cdot IDCG(D)},\\
\end{split}
\end{equation}
where $discount(d)$ is the discount for document $d$ in NDCG, IDCG is the ideal DCG, and $\sigma$ is set to 2 for avoiding gradient vanishing in LightGBM for (\ref{P_t(d)}). Note that uMart does not need gradients of the loss function and thus the challenge of (\ref{window_P_t(d)}) not being differentiable does not apply to this model. It is easy to see that uMart is a generalization of LambdaMart, and uMart becomes LambdaMart (a pairwise model) when the window size is 1. The contribution document $d$ makes to the second derivative is

\begin{equation}\label{second_derivative}
\begin{split}
&\frac{\partial L^2(F_m; R, \{d\})}{\partial^2 F_m}\\
&= 2 \sigma^2 \sum_{d' \in \tilde{s}_{r(d)}}  \Delta Z(d, d') \frac{e^{\sigma \cdot F_m(d')}}{e^{\sigma \cdot F_m(d)} + k(\tilde{s}_{r(d)})} \\
&- 2 \sigma^2 \sum_{\hat{d} \in D \setminus s_{r(d)}} \Delta Z(\hat{d},d) \\
& \frac{e^{\sigma \cdot F_m(d)}}{k(\tilde{s}_{r(\hat{d})}) + e^{\sigma \cdot F_m(\hat{d})}} (1 - \frac{e^{\sigma \cdot F_m(d)}}{k(\tilde{s}_{r(\hat{d})}) + e^{\sigma \cdot F_m(\hat{d})}})\\
\end{split}
\end{equation}

\subsection{urBoost Model}
\label{urBoost}
The classical Plackett-Luce model assumes that the process of sampling documents at different steps is independent. Therefore, documents share the same global scoring function in uRank as well as other listwise models. However, this is a questionable assumption. Documents we can select from $s_t$ itself depend on $s_1 ... s_{t-1}$. In this section, we introduce our fourth model urBoost that adapts RNN functions for modeling dependencies from documents ranked at previous steps.

We assume the prediction score of a document $d$ in $s_t$ comes from two parts: (1) individual relevancy that is directly captured in query-document features, (2) conditional relevancy given documents ranked at previous steps $1,...,t-1$. Similar to uRank, we use three fully-connected neural network layers that take the raw query-document features $X$ as input to generate feature embedding matrix $\tilde{X}$. $\tilde{X}$ is the output of the second fully-connected layer. Let function $\tilde{x}(d)$ return the feature embeddings of document $d$, which does not change for the same document in different sets.

In the training phase, we apply the standard RNN function to each of the documents in set $s_t$ to obtain their hidden embeddings $\{\tilde{h}_t(d)\}_d$. The hidden embeddings are used to calculate the conditional prediction scores. Function $f(d)$ returns the conditional prediction score of document $d$ in set $s_t$, and we use (\ref{loss}) and (\ref{gradients_full}) to calculate the loss and residuals. The formulas for calculating $\tilde{h}_t(d)$, $f(d)$ for $d \in s_t$, and $h_t$ are
\begin{equation} \label{urBoost_eq}
\begin{split}
\tilde{h}_t(d) &= rnn(\tilde{x}(d), h_{t-1}, \tilde{W}), \\
f(d) &= [\tilde{h}_t(d), \tilde{x}(d)] w, \\
h_t &= \text{pooling } \{\tilde{h}_t(d) | d \in c_t\},
\end{split}
\end{equation}
\noindent where $w$ and $\tilde{W}$ are  trainable.

\begin{table*}[t]
\centering
\scalebox{0.9}{
\begin{tabular}{|l|r|r|r|r|r|r|r|}
\hline
& Queries 	& Documents 	& Features 	& Rating levels 	& LR (\%) & MPQ		& ALRPQ (\%)   \\ \hline
OSHUMED       & 106                          & 16,140                    	& 45                             & 3       & 68.12 & 320 & 71.87                           \\ \hline
MQ2007        & 1692                         & 69,623                    	& 46                             & 3       & 87.22 & 147 & 73.74                           \\ \hline
MSLR-WEB10K & 10,000                       & 1,200,192                    & 136                            & 5       & 51.75 & 908 & 56.02                           \\ \hline
MSLR-WEB30K & 31,531                       & 3,711,000                    & 136                            & 5       & 52.22 & 1251 & 53.33                           \\ \hline
\end{tabular}
}
\caption{The columns correspond to the number of queries, the number of documents, the number of features, the rating levels, the ratio of the number of documents having the lowest rating over the total number of documents (LR), the maximum number of documents per query (MPQ), and the average ratio of the number of documents having the lowest rating per query over the number of documents per query (ALRPQ)}
\label{datasets}
\end{table*}

We have attempted max-pooling and average-pooling for calculating $h_t$, and max-pooling performs slightly better. Besides, we have also tried replacing RNN with an LSTM or GRU, but the simple RNN cell works best in the experiments. Each and every sequence is very short, i.e., the maximum number of unique ratings is 5, thus an RNN is sufficient for capturing dependencies in such short sequences. Also, an RNN has fewer weight matrices to train. Although it is common to use a trained RNN to predict outputs one by one during inference, we have found out that using $f(d)$, $d \in s_1$ during inference works better in both NDCG performances and inference time complexity. This is due to the fact that during inference if an earlier position makes a wrong prediction about a ranking level, this has a bad effect on later ranking levels. Hence, we only use $f(d)$, $d \in s_1$ for sorting in the inference phase.

Ranking refinement models such as DLCM \cite{ai2018dlcm} needs two training phases and its performance relies on its base model. Moreover, the increased inference time due to predicting documents one by one limits its usage in real-world applications. In each step, different from DLCM that feeds a single document to an RNN cell, urBoost feeds a set of documents to RNN cells. The inference of urBoost only takes the conditional prediction scores at the first step; thus it does not increase inference time. 

\section{Experiments}
\label{Experiments}

\subsection{Datasets}
We use four widely used learning-to-rank datasets OSHUMED \cite{qin2010letor}, MQ2007, MSLR-WEB10K, and MSLR-WEB30K \cite{qin2013introducing} that come with train/validation/test data. Each dataset contains five folds and has multi-level ratings ranked by domain experts. The statistics of the datasets are shown in Table \ref{datasets}. Especially from the LR and ALRPQ ratios, we can see that the majority of documents in the datasets have the lowest ratings. These are the documents that our models do not need to process as they are in the final set $s_{|R|}$. Also, the models decrease the maximum number of steps that a strict Plackett-Luce model (e.g., ListMLE, p-ListMlE) needs to go through from the MPQ numbers to rating levels in the table.

\begin{table*}[t]
\centering
\begin{tabular}{|l|r|r|r|r|r|r|r|r|}
\hline
\multicolumn{9}{|l|}{OSHUMED}                                                                           \\ \hline
          & NDCG@1 	& ERR@1 		 & NDCG@3   & ERR@3      & NDCG@5   & ERR@5        & NDCG@10  & ERR@10\\ \hline
uRank 	     & * 0.602  & * 0.449  & * 0.511  & * 0.547  & * 0.481  & * 0.573  & * 0.455  & * 0.586 \\ \hline
uBoost       & * 0.609  & * 0.454  & * \textbf{0.531}  & * \textbf{0.555}  & * 0.493  & * \textbf{0.580}  & * 0.464  & * \textbf{0.594} \\ \hline
uMart 	     & 0.530  & 0.395  & 0.452  & 0.497  & 0.431  & 0.523  & 0.411  & 0.538 \\ \hline
urBoost      & * \textbf{0.612}  & * \textbf{0.457}  & * 0.529  & * \textbf{0.555} & * \textbf{0.494}  & * 0.578  & * \textbf{0.467}  & *0.590 \\ \hline
MDPRank      & 0.500  & 0.317  & 0.440  & 0.415  & 0.421  & 0.449  & 0.409  & 0.464 \\ \hline
LambdaMart-L & 0.506  & 0.375  & 0.445  & 0.488  & 0.426  & 0.514  & 0.409  & 0.527  \\ \hline
LambdaMart-R & 0.468  & 0.346  & 0.412  & 0.463  & 0.389  & 0.485  & 0.380  & 0.500   \\ \hline
LambdaLoss   & 0.563  & 0.389  & 0.449  & 0.489  & 0.441  & 0.518  & 0.420  & 0.531	  \\ \hline
p-ListMLE    & 0.469  & 0.311  & 0.410  & 0.430  & 0.398  & 0.453  & 0.396  & 0.472	   \\ \hline
\end{tabular}
\caption{NDCG and ERR performance on the OSHUMED dataset}
\label{OSHUMED NDCG}
\end{table*}


\begin{table*}[t]
\centering
\begin{tabular}{|l|r|r|r|r|r|r|r|r|}
\hline
\multicolumn{9}{|l|}{MQ2007}                                                                            \\ \hline
        & NDCG@1 	& ERR@1 		 & NDCG@3   & ERR@3      & NDCG@5   & ERR@5        & NDCG@10  & ERR@10\\ \hline
uRank        & * 0.481  & * 0.295  & * 0.482  & * 0.404  & * 0.490  & * 0.430  & * 0.526  & * 0.448	\\ \hline
uBoost       & * 0.489  & * 0.300  & * 0.481  & * 0.406  & * 0.491  & * 0.433  & * 0.523  & * 0.450 \\ \hline
uMart        & *0.480  & *0.295  & *0.477  & *0.404  & *0.487  & *0.432  & 0.519  & *0.449 \\ \hline
urBoost      & * \textbf{0.492}  & * \textbf{0.301}  & * \textbf{0.486}  & *\textbf{0.410}  & * \textbf{0.493} & * \textbf{0.436} &  * \textbf{0.524} & *\textbf{0.453} 		 \\ \hline
MDPRank  	 & 0.406  & 0.166  & 0.410  & 0.257  & 0.417  & 0.286  & 0.442  & 0.309  \\ \hline
LambdaMart-L & 0.469  & 0.234  & 0.472  & 0.346  & 0.478  & 0.362  & 0.512  & 0.369 \\ \hline
LambdaMart-R & 0.465  & 0.224  & 0.464  & 0.315  & 0.474  & 0.341  & 0.508  & 0.360	\\ \hline
LambdaLoss   & 0.406  & 0.192  & 0.412  & 0.274  & 0.419  & 0.296  & 0.448  & 0.313	 \\ \hline
p-ListMLE    & 0.380  & 0.177  & 0.338  & 0.236  & 0.387  & 0.255  & 0.372  & 0.272  \\ \hline
\end{tabular}
\caption{NDCG and ERR performance on the MQ2007 dataset}
\label{MQ2007 NDCG}
\end{table*}


\begin{table*}[t]
\centering
\begin{tabular}{|l|r|r|r|r|r|r|r|r|}
\hline
\multicolumn{9}{|l|}{MSLR-WEB10K}  \\ \hline
             & NDCG@1 	& ERR@1 	& NDCG@3   & ERR@3  & NDCG@5   & ERR@5  & NDCG@10  & ERR@10  \\\hline
uRank 		  & 0.458  & 0.387  & 0.441  & 0.501  & 0.443  & 0.526  & 0.458  & 0.542   \\ \hline
uBoost 		  & 0.459  & 0.388  & 0.441  & 0.502  & 0.444  & 0.528  & 0.460  & 0.544  \\\hline
uMart 		  & * \textbf{0.481}  & * \textbf{0.407}  & * \textbf{0.468}  & * \textbf{0.525}  & * \textbf{0.474}  & * \textbf{0.550}  & * \textbf{0.495}  & * \textbf{0.565} \\ \hline
urBoost 	  	 & 0.461  & 0.390  & 0.442  & 0.503  & 0.444  & 0.529  & 0.462  & 0.545 	\\\hline
LambdaMart-L  & 0.468  & 0.396  & 0.451  & 0.512  & 0.457  & 0.538  & 0.475  & 0.553 \\\hline
LambdaMart-R  & 0.421   & 0.326	  & 0.419 	& 0.400	 & 0.427  & 0.325  & 0.446  & 0.344   \\\hline
LambdaLoss    & 0.423  	& 0.308   & 0.423  	& 0.378  & 0.427  & 0.294  & 0.446  & 0.322	\\\hline
p-ListMLE     & 0.407   & 0.286   & 0.396   & 0.370  & 0.377  & 0.274  & 0.420  & 0.300\\ \hline
\end{tabular}
\caption{NDCG and ERR performance on the MSLR-WEB10K dataset}
\label{MSLR-WEB10K NDCG}
\end{table*}


\begin{table*}[t]
\centering
\begin{tabular}{|l|r|r|r|r|r|r|r|r|}
\hline
\multicolumn{9}{|l|}{MSLR-WEB30K}  \\ \hline
        & NDCG@1 	 & ERR@1 	& NDCG@3   & ERR@3   & NDCG@5   & ERR@5   & NDCG@10  & ERR@10   \\ \hline
uRank 		   & 0.465  & 0.401   & 0.450    &0.515    &0.453  &0.517 & 0.471   & 0.545		\\ \hline
uBoost  	       & 0.474   &0.403   & 0.452    & 0.517  & 0.457  &0.543 & 0.472    & 0.548		\\ \hline
uMart 		   & * \textbf{0.505}  & * \textbf{0.429}  & * \textbf{0.490}  & * \textbf{0.545}  & * \textbf{0.494}  & * \textbf{0.570}  & * \textbf{0.512}  & * \textbf{0.584} \\ \hline
urBoost 	       & 0.478  & 0.405  & 0.455  & 0.518  & 0.458  & 0.543  & 0.472  & 0.558 	\\\hline
LambdaMart-L   & 0.489  & 0.415  & 0.468  & 0.530  & 0.472  & 0.556  & 0.490  & 0.571   \\\hline
LambdaMart-R   & 0.458  & 0.353	 & 0.442  & 0.315  & 0.448 	& 0.336  & 0.464  & 0.440   \\\hline
LambdaLoss     & 0.430  & 0.310  & 0.399  & 0.307  & 0.410  & 0.400  & 0.428  & 0.377	 	\\\hline
p-ListMLE      & 0.411  & 0.310  & 0.386  & 0.285  & 0.394  & 0.347  & 0.410  & 0.384    \\\hline
\end{tabular}
\caption{NDCG and ERR performance on the MSLR-WEB30K dataset}
\label{MSLR-WEB30K NDCG}
\end{table*}

\subsection{Experimental Settings}
\label{Experimental Settings}
For neural network models, we consider documents that are in the same query to be in the same batch for training, validation, and test phases. More specifically, each batch contains all the documents of a single query. Hence, the actual batch sizes vary in different queries. We also perform feature normalization using the min and max values within each batch, instead of normalizing among all the documents. We notice this feature normalization method can enhance all models compared to the latter. 

The two hidden fully-connected layers in uRank, uBoost, and urBoost have dimensions [100, 50] for the OSHUMED dataset, dimensions [100, 100] for the MQ2007 dataset, and dimensions [200, 200] for the MSLR datasets. We use the ELU function as the activation function for the input layer and the two hidden layers. The corresponding hidden fully-connected layers for their residual calculation have the same dimension size setting. We use the Adam optimizer with the initial learning rate $10^{-4}$ for the OSHUMED dataset, and the initial learning rate is $10^{-5}$ for the other datasets in training. The weights that yield the highest NDCG@1, then the highest NDCG@3, NDCG@5, and NDCG@10 on a validation set are used for the corresponding test. More specifically, we update weights: (1) when a higher NDCG@1 is found, or (2) when an incoming NDCG@1 is the same as the current highest NDCG@1 up to tolerance $10^{-6}$, but a higher NDCG@3 is found, etc. We have found out this weight selection strategy works better than strategies that use a single metric such as the highest validation NDCG@1 or the lowest validation loss. A model stops training when no better weights are observed on validation data in consecutive 200 epochs. We clip the global gradient norm with 5 for neural network models before gradient updates. For the uBoost and urBoost models, the maximum number of weak learners is 5, or it stops when there is no NDCG improvement. We have attempted different coefficients for weak-learners, and constant 1 works well for uBoost and urBoost. As random access and partitioning are expensive operations in Tensorflow, (\ref{window_P_t(d)}) is only implemented in the uMart model and all other models use (\ref{P_t(d)}). The window size $u$ in the experiments for the OSHUMED, MSLR-WEB10k, and MSLR-WEB30k datasets is 2, and the window size for the MQ2007 dataset is 10. Further enlarging the window sizes does not significantly improve NDCG but slightly increases training time. The learning rate for the Mart-based models is 0.05.

We have noticed that widely-used open-source repositories for state-of-the-art ranking models have slightly different NDCG settings, especially regarding documents with rating 0. If all the documents in a query have the lowest rating of 0, the IDCG value at any position is 0. Some open-source repositories take the NDCG value at any position as 0, some take it as 1, and some take it as 0.5. If the number of queries with all 0 labels is large (e.g., MQ2007 and MSLR datasets), the NDCG design choices can noticeably influence average NDCG results. In order to avoid confusion and to have a fair comparison, we leave out queries with all 0 labels for the NDCG calculation of all models and ranking positions in our experiments. For each fold, we use the weights that have the best NDCG performance on the validation dataset to evaluate the test dataset. We publish the code of our neural network models implemented in Tensorflow, the code of uMart (modified based on LightGBM), and parameter details on GitHub\footnote{https://github.com/XiaofengZhu/uRank\_uMart}.

\begin{figure*}[pt]
\begin{subfigure}[b]{0.24\textwidth}
  \includegraphics[width = \textwidth]{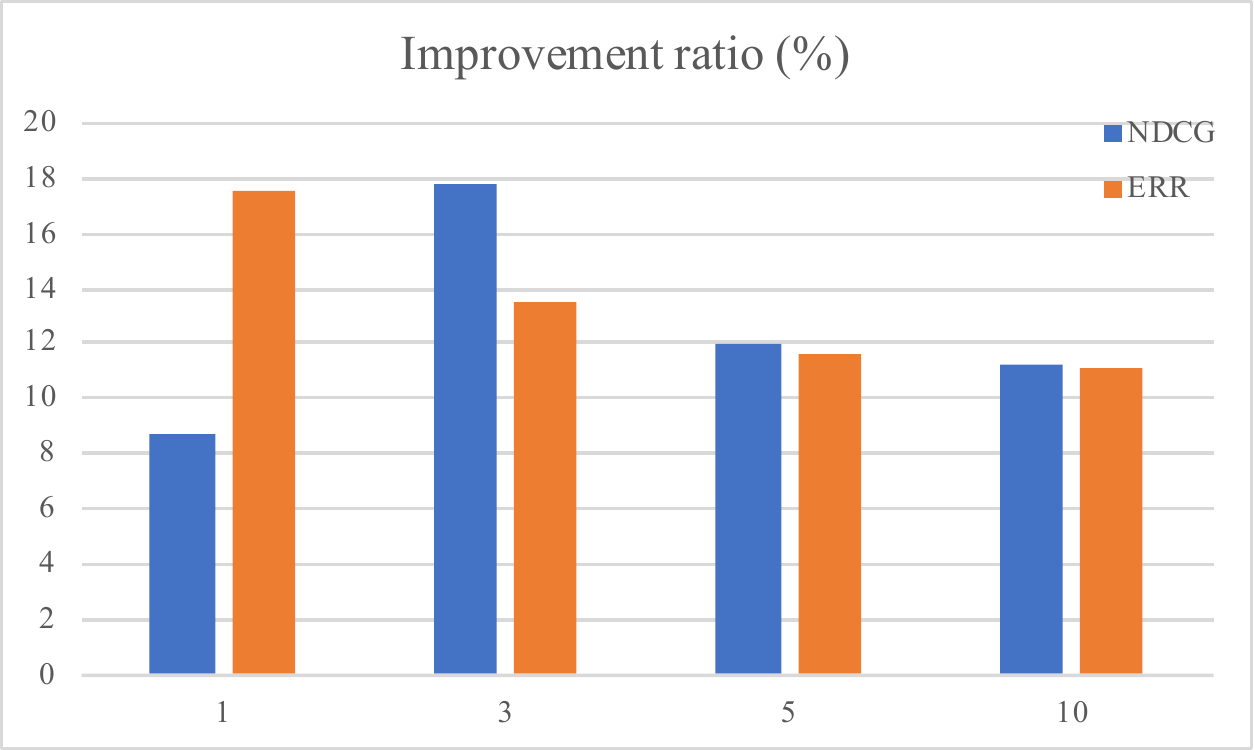}
  \caption{urBoost over LambdaLoss\\on the OHSUMED dataset}
  \label{fig:OHSUMED}
\end{subfigure}%
\begin{subfigure}[b]{0.24\textwidth}
  \includegraphics[width =\textwidth]{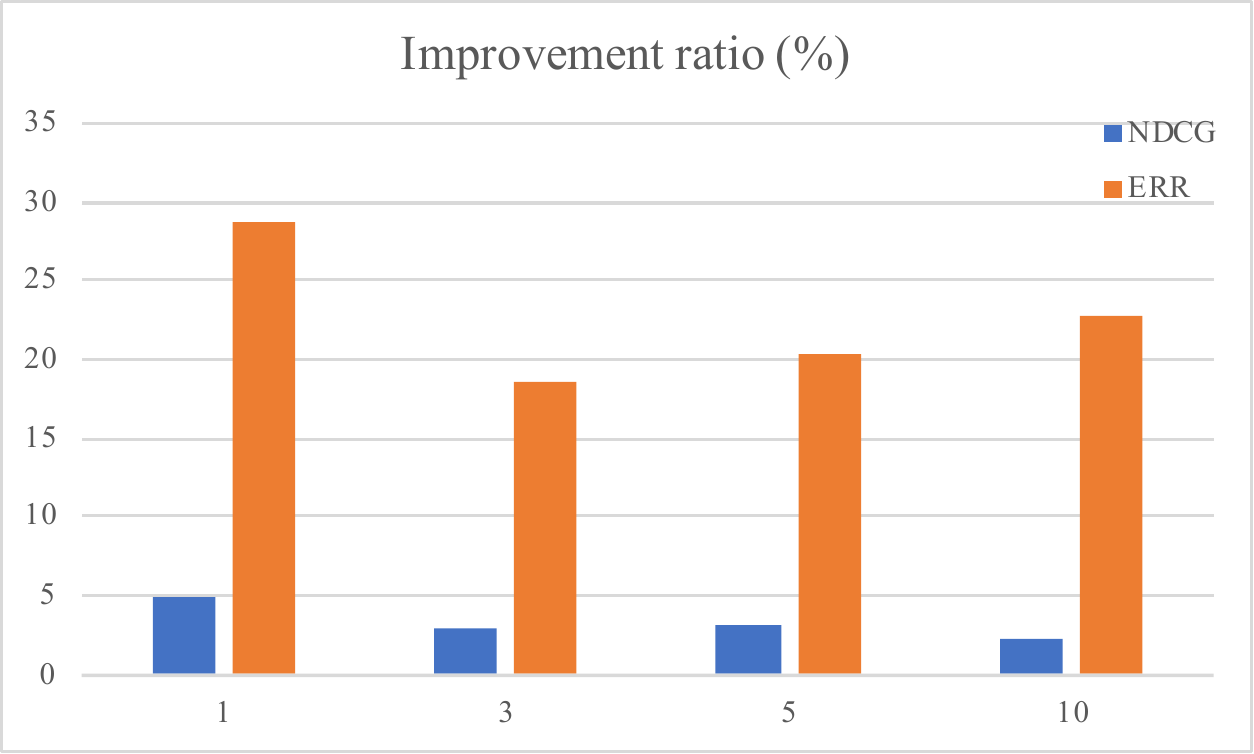}
  \caption{urBoost over LambdaMart-L\\on the MQ2007 dataset}
  \label{fig:MQ2007}
\end{subfigure}%
\begin{subfigure}[b]{0.24\textwidth}
  \includegraphics[width = \textwidth]{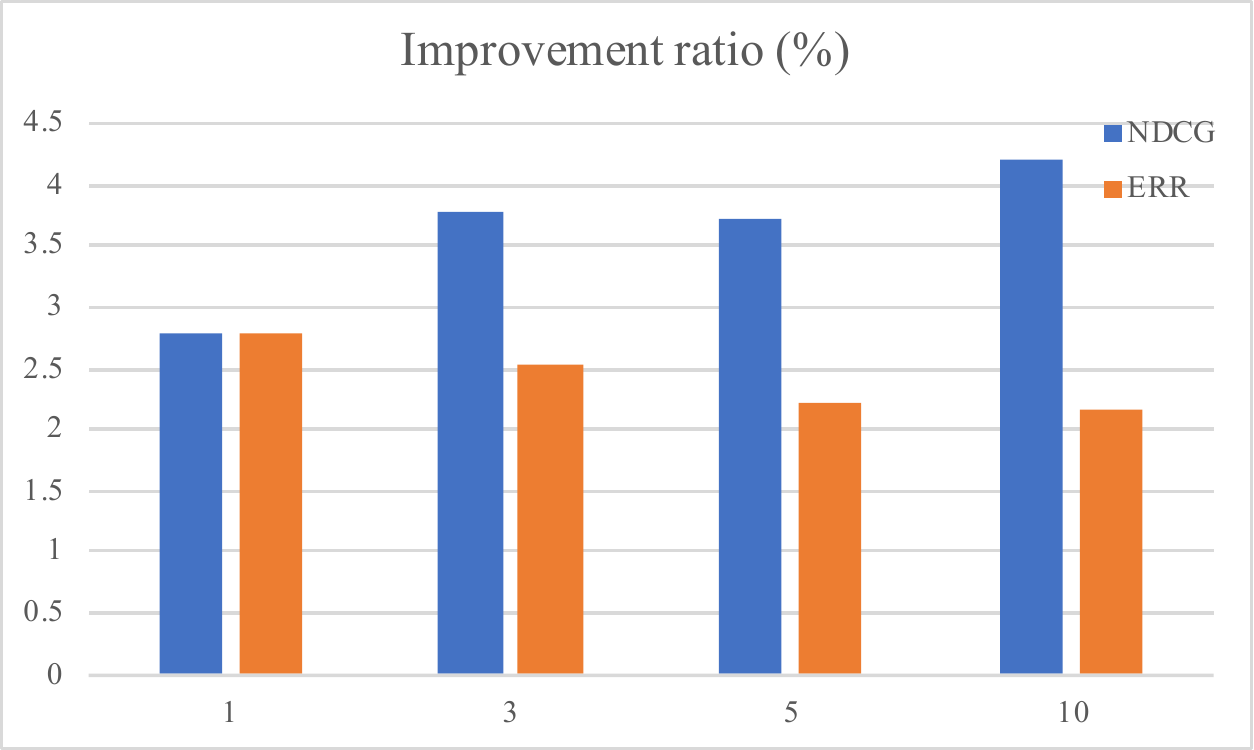}
  \caption{uMart over LambdaMart-L\\on the MSLR-WEB10K dataset}
  \label{fig:MSLR-WEB10K}
\end{subfigure}%
\begin{subfigure}[b]{0.24\textwidth}
  \includegraphics[width =\textwidth]{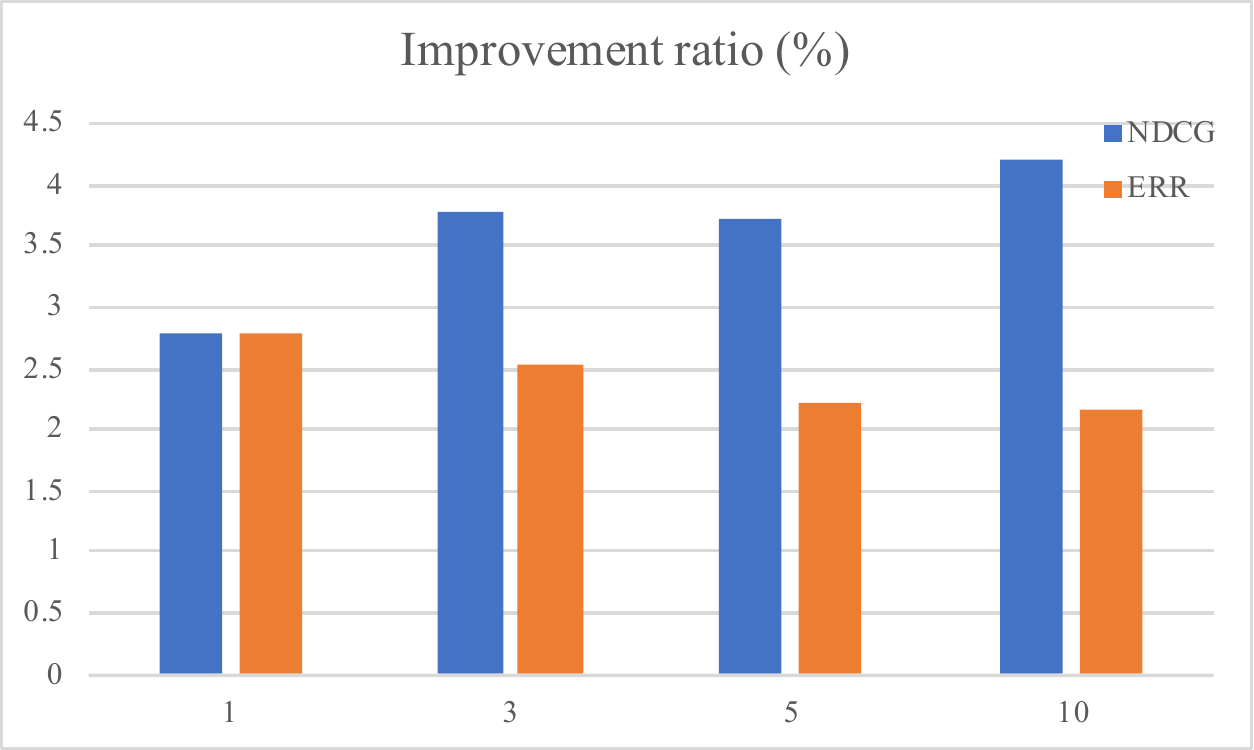}
  \caption{uMart over LambdaMart-L\\on the MSLR-WEB30K dataset}
  \label{fig:MSLR-WEB30K}
\end{subfigure}
\caption{Relative NDCG and ERR improvements}
\label{Relative NDCG and ERR improvements}
\end{figure*}

\begin{table*}[t]
\centering
\scalebox{0.9}{
\begin{tabular}{|l|r|r|r|r|}
\textbf{MSLR-WEB10K}       & NDCG@1 & NDCG@3  & NDCG@5  & NDCG@10     \\ \hline
uMart                   		&\textbf{0.496} & \textbf{0.484} & \textbf{0.490} & \textbf{0.509}    \\ \hline
LambdaMart-L                 & 0.484 & 0.478 & 0.484 & 0.504     \\ \hline
\hline
\textbf{MSLR-WEB30K}       & NDCG@1 & NDCG@3  & NDCG@5  & NDCG@10     \\ \hline
uMart                   	 	& \textbf{0.522} & \textbf{0.507} & \textbf{0.510} & \textbf{0.529}    \\ \hline                    		
LambdaMart-L                 & 0.518 & 0.504 & 0.509 & 0.527     \\ \hline
\end{tabular}
}
\caption{LightGBM evaluation on the MSLR-WEB10K and MSLR-WEB30k datasets}
\label{LigthGBM NDCG}
\end{table*}

\section{Results}
\label{Results}

We ran all experiments using five-fold cross-validation. We report the model performance using NDCG and ERR at positions 1, 3, 5, and 10 respectively, which are metrics designed for multi-level ratings. We compare our models to four other models in recent publications: LambaMart, p-ListMLE \cite{lan2014position}, MDP Rank, and LambdaLoss \cite{wang2018lambdaloss}. We chose p-ListMLE because it was reported to be the best Plackett-Luce model \cite{lan2014position}. We only report the results of MDP on OSHUMED and MQ2007 datasets because it leads to Out-Of-Memory on MSLR datasets.
For LambdaMart, we use the implementation in RankLib\footnote{https://sourceforge.net/p/lemur/wiki/RankLib/}, denoted as LambdaMart-R in Tables \ref{OSHUMED NDCG} to \ref{MSLR-WEB30K NDCG}, and the implementations in LightGBM\footnote{https://github.com/microsoft/LightGBM} \cite{meng2016communication, ke2017lightgbm}, denoted as LambdaMart-L. We also include the results of LambdaLoss and p-ListMLE supported in the TF-Ranking framework \cite{TensorflowRanking2018}. LambdaLoss is the ensemble model that combines NDCG LOSS++ \cite{wang2018lambdaloss} and group-wise functions \cite{ai2018learninggroupwise}. TF-Ranking takes a list of documents of the same size from each query by default. We use the maximum list size in a dataset -- the MPQ numbers in Table \ref{datasets} for LambdaLoss and p-ListMLE in the experiments. We also use the same gradient norm cut-off value of 5 for LambdaLoss and p-ListMLE.

Tables \ref{OSHUMED NDCG} to \ref{MSLR-WEB30K NDCG} show the NDCG and ERR performance of our models and the benchmark models on the four datasets, respectively.  Boldface indicates the highest score in each column, and * denotes a significant improvement over LambdaMart-L based on the t-test at level of 0.05 for datasets MQ2007 and the two MSLR datasets, and over LambdaLoss for OHSUMED (these are the best benchmark models for the corresponding datasets). The relative improvements of our best models over the best benchmark models in the four datasets are shown in Figure \ref{Relative NDCG and ERR improvements}. 

uBoost and urBoost have the best performance on the OSHUMED and MQ2007 datasets. uMart has the best performance on the MSLR-WEB10K and MSLR-WEB30K datasets. Mart-based models generally perform better than neural networks on the MSLR datasets. Our intuition is that most features in the MSLR datasets are sparse, which favors Mart-based models. Mart-based models use 200-1000 trees/learners with 31-255 leaves per tree and 255-10,000 bins. Compared to uBoost, urBoost has an additional RNN network, but urBoost can use smaller fully-connected layer sizes and the smallest number of weak learners (1 or 2 learners) to perform well.




For fairness, we include the NDCG results shown in Table \ref{LigthGBM NDCG} using the evaluation method in LightGBM, which is validating and testing directly on a test dataset and NDCG being 1 at any position if all labels are 0s \footnote{This setting of LightGBM was for a fair comparison with earlier gradient boosting models such as XGBoost \cite{chen2016xgboost}.}. The hyper-parameters we use come from the original paper, and the NDCG performance of LightGBM-L in this experiment matches with the authors' report. Model uMart also outperforms the benchmark for this experimental setup.

All neural network models are trained and tested on a single GeForce TITAN X GPU card with 12 GB memory. For the Mart-based models, i.e. uMart, and LambdaMart, we use 16 CPUs. We have optimized the hyper-parameters in all benchmark models. By utilizing unique ratings, the uRank model is 7-14 times faster than p-ListMLE and 15-32 times faster than MDP Rank depending on the underlying dataset. The tensor-based algorithms speed up training by 2-3 times in our experiments compared to implementing uRank, uBoost, and urBoost models using "for loops." We do not compare the training time of the neural network models and the Mart-based models as they do not have the same computational resource.

%

\section{Conclusion}
\label{Conclusion}
In this paper, we propose a new loss function for multi-level ratings and associated listwise learning-to-rank models uRank, uBoost, uMart, and urBoost using efficient matrix calculations. The models overcome the tie issue of existing Plackett-Luce models by maximizing the likelihood of selecting documents with high ratings over documents with low ratings. Furthermore, urBoost provides dependencies from ranked documents using RNN functions, which enhances the performance. When compared to Mart-based models, we conclude that Mart-based models have better performance compared to neural networks especially when features are sparse. However, the Mart-based models consume more memory. Nevertheless, uRank, uBoost, and urBoost models would have an advantage over Mart-based models by seamlessly linking to diverse deep learning embedding techniques when plain text features are present \cite{haldar2019applying, li2019combining}. 
An interesting future direction would be leveraging the strengths of neural networks and trees. Another future study would be exploring more ways of capturing ranking dependencies and studying relationships among similar queries.

\bibliographystyle{ACM-Reference-Format}
\bibliography{ref}

\end{document}